\documentclass[11pt]{article}

\usepackage[latin1]{inputenc}
\usepackage[T1]{fontenc}
\usepackage[english]{babel}
\usepackage{amsfonts}
\usepackage{amssymb}
\usepackage{amsmath}
\usepackage{amsthm}
\usepackage{graphicx}

\newtheorem{proposizione}{Proposition}
\newtheorem{lemma}{Lemma}
\newtheorem{teorema}{Theorem}
\newtheorem{corollario}{Corollary}
\newtheorem{remark}{Remark}
\newcommand{\meanv}[1]{\left\langle#1\right\rangle}
\newcommand{\arctanh}[1]{\mbox{arc}\tanh #1}

\begin{document}
\title{A mechanical approach to mean field spin models}
\author{Giuseppe Genovese \footnote{Dipartimento di Fisica,
    Sapienza Universit\`a di Roma} \ and Adriano Barra$^*$ \footnote{Dipartimento di Matematica,
    Universit\`a  di Bologna}}
\date{December, 11, 2008}
\maketitle

\begin{abstract}
Inspired by the bridge pioneered by Guerra among statistical
mechanics and analytical mechanics on $1+1$ continuous
Euclidean space-time, we built a self-consistent method to solve
for the thermodynamics of mean-field models,
whose order parameters self average. We show the whole procedure
by analyzing in full details the simplest test case, namely the
Curie-Weiss model. Further we report some applications also to
models whose order parameters do not self-average, by using the
Sherrington-Kirkpatrick spin glass as a guide.
\end{abstract}

\section*{Introduction\\}

Mean field statistical mechanics of discrete systems is
experiencing a massive increasing of interest for several reasons.
Born as an infinite dimensional limit of a theoretical background
for condensed matter physics, mean field statistical mechanics
become immediately appealing for its possibility of being solved
(even though this happens exactly for really a few models
\cite{talabook}), still retaining several features of more
realistic systems with finite dimensionality.

Furthermore, and maybe nowadays, foremost, its range of
applicability is continuously spreading such that, so far, it is
one of the key tools for the investigation of several models far
away from physics like biological or social networks (see for
instance, respectively, \cite{immune}\cite{parnet} and \cite{noi}
\cite{watts}): all systems where the mean field nature of the
description is not a limitation and whose rigorous or heuristic
analysis was, in past decades, unimaginable.

As a consequence the need for methods in statistical mechanics is
one of the fundamental enquiries raised to theoretical physicists
and mathematicians involved in the field.

In this paper, inspired by a pioneering work of Francesco Guerra
\cite{sum-rules}, we develop an alternative approach to standard
statistical mechanics to solve for the thermodynamics of systems
whose order parameters self-average.

With the aim of presenting the theory also to readers who may not
be experts in statistical mechanics, we apply our scheme to the
simplest and most well known Curie-Weiss (CW) model, which we
solve in full detail, for the sake of simplicity, linking our
procedures with general statistical mechanics models via frequent
remarks spread throughout the whole paper.

As the largest interest is payed to complex systems, after the CW,
we analyze the Sherrington-Kirkpatrick (SK) model, in the replica
symmetric regime, subjected to an external field.

\section*{Guerra Action for mean field spin models}

Even though we will be interested in observable's behavior  once
the thermodynamic limit is taken, let us consider a large set of
$N$ Ising spins $\sigma_i = \pm 1$, $i \in (1,...,N)$. Let us deal
with a generic mean-field spin model, described by the Hamiltonian
\begin{equation}\label{eq:ISINGmean-field_H}
H_N(\sigma)=-\sum_{(i,j)}^N \chi_{ij} \sigma_{i}\sigma_{j},
\end{equation}
where $\chi_{ij}$ is a two body interaction matrix. The main
quantity of interest in statistical mechanics is the infinite
volume limit of the free energy $f(\beta)= \lim_{N \rightarrow
\infty}f_N(\beta) = \lim_{N \rightarrow \infty}
-\beta^{-1}A_N(\beta)$, where $A_N(\beta)$ is the pressure and is
related to the Hamiltonian via
$$
A_N(\beta)= \frac{1}{N}\ln \sum_{\sigma} \exp(-\beta
H_N(\sigma)).
$$

We stress here (even though we will not deal with disordered
systems in the first part of the work) that for the SK model it is
usually expected to consider the quenched average of the free
energy \cite{sum-rules}, however, without explicit expectation
over the random coupling we mean its value $\chi$-almost surely in
the sense of the first Borel-Cantelli lemma.
\newline
It is useful to consider the one body interaction, of the same
nature of Hamiltonian, that we call \textit{cavity field}
$$
H^{\prime}_N(\sigma)=-\sum_i^N \chi_i \sigma_i.
$$
We define further a two parameters \textit{Boltzmannfaktor} $B(x,t)$ and a
relative Gibbs measure $\langle . \rangle_{(x,t)}$ as:
\begin{eqnarray}
B_N(x,t)&=&\exp\left(\theta(t)H_N+\theta(x)H^{\prime}_N\right),\label{eq:Ising-B}\\
\meanv{f(\sigma)}_{(x,t)}&=& \ \
\frac{\sum_{\sigma}f(\sigma)(B(x,t))}{\sum_{\sigma}(B(x,t))}\label{eq:Ising-misura},
\end{eqnarray}
where $\theta$ is a increasing function, vanishing at the origin,
strictly dependent by the form of interaction. Eventually a
magnetic field can be added in (\ref{eq:ISINGmean-field_H}), and
therefore in (\ref{eq:Ising-B},\ref{eq:Ising-misura}).

We define the Guerra action $\varphi(x,t)$ for a mean field model as the solution of the Hamilton-Jacobi differential equation
\begin{equation}\label{eq:GA}
\partial_t \varphi_N (x,t)+\frac{1}{2}\left(\partial_x \varphi_N (x,t)\right)+V_N(x,t)=0,
\end{equation}
with suitable boundary condition.

Furthermore the function $u(x,t)=\partial_x \varphi (x,t)$
satisfies
\begin{equation}\label{eq:Gu}
\partial_t u_N (x,t)+u_N(x,t)\partial_x u_N (x,t)+\partial_x V_N(x,t)=0.
\end{equation}
The Guerra action $\varphi_N(x,t)$ is related to the pressure of
the model $A_N(\sigma)$, in a way that will be specified later,
case by case.

Consequently even the  potential function $V_N(x,t)$ expresses
thermodynamical quantities of the case study (i.e. in CW and SK
models we investigate, it turns out to be the self-averaging of
the order parameters).

We will be interested throughout the paper in the region where
$V(x,t)=0$, when we can always solve our equations \footnote{We
stress however that the formalism we develop can still be applied
to general constrained problems ($V(x,t) \neq 0$), even though
their resolutions can be prohibitive}. In fact these problems are
largely studied in literature far away from statistical mechanics
\cite{evans}. In particular some Theorems, due to Lax \cite{lax},
are helpful, since under certain hypothesis (that in a nutshell
are related to the uniform convexity of the quantity
$\frac{1}{2}(\partial_x\varphi(x,t))^2$), give the form
of the unique solution of (\ref{eq:GA}) and (\ref{eq:Gu}) (related
by a derivation). Following Lax we can state the next
\begin{teorema}\label{th:lax}
For a general differential problem
\begin{equation}\label{eq:HJ-TL-free}
\left\{
\begin{array}{rclll}
&& \partial_t\varphi(x,t)+\frac{1}{2}(\partial_x\varphi(x,t))^2= 0 &\:&\mbox{in }{\mathbb R}\times(0,+\infty)\\
&& \varphi(x,0)=h(x)&\:&\mbox{on}{\mathbb R}\times \{t=0\},
\end{array}
\right.
\end{equation}
and
\begin{equation}\label{eq:burgers-TL-free}
\left\{
\begin{array}{rclll}
&& \partial_t u(x,t)+u(x,t)\partial_x u(x,t)= 0&\:&\mbox{in }{\mathbb R}\times(0,+\infty)\\
&& u(x,0) = g(x)&\:&\mbox{on }{\mathbb R}\times \{t=0\},
\end{array}
\right.
\end{equation}
where $h(x)$ is Lipschitz-continous, and
$g(x)=h^{\prime}(x)\in\mathcal{L}^{\infty}$, it does exist and it
is unique the function $y(x,t):
\mathbb{R}\times\mathbb{R}^+\to\mathbb{R}$ such that
\begin{eqnarray}\label{eq:phi}
\varphi(x,t)&=&\min_y\left\{ \frac{t}{2}\left(\frac{x-y}{t}\right)^2+h(y) \right\}\nonumber\\
&=&\frac{t}{2}\left(\frac{x-y(x,t)}{t}\right)^2 + h(y(x,t))
\end{eqnarray}
is the unique weak solution of (\ref{eq:HJ-TL-free}), and
\begin{equation}\label{eq:u}
u(x,t)=\frac{x-y(x,t)}{t}
\end{equation}
is the unique weak solution of (\ref{eq:burgers-TL-free}). Furthermore, the function $x\to y(x,t)$ is not-decreasing.
\end{teorema}

It is worthwhile to remark that the choice of looking for weak
solution (that arises naturally in the Lax's theorems) may look as
redundant in our case, since we deal with physical quantities (in
general smooth functions). Actually it prevents us from the
eventual discontinuities of the solutions of (\ref{eq:HJ-TL-free})
and (\ref{eq:burgers-TL-free}). However, a strong solution is a
weak solution too and there is no need to change the essence of
the Theorem.

Let us start applying this framework to the CW model.

\section*{Mean field ferromagnet as a 1-dimensional fluid}

The mean field ferromagnetic model is defined by the Hamiltonian
$$
H_N (\sigma) =-\frac{1}{N}\sum_{(i,j)}^N
\sigma_i\sigma_j+h\sum_i^N \sigma_i.
$$
It is easily seen that we have resumed in the Hamiltonian both the
two body and one body interaction\footnote{In ferromagnet the
cavity field coincides with the external field}. Thus, choosing
$\theta(a)=a$, we can write the $(x,t)$-dependent
\textit{Boltzmannfaktor} as
$$
B_N(x,t)=\exp\left(
\frac{t}{N}\sum_{(i,j)}^N\sigma_i\sigma_j+x\sum_i^N\sigma_i
\right).
$$
\begin{remark}
When dealing with the ferromagnetic Boltzmannfaktor $B_N(x,t$)
above, classical statistical mechanics is recovered of course, in
the free field case, by setting $t=\beta$ and $x=0$.

In the same way the averages $\langle . \rangle_{(x,t)}$ will be
denoted by $\langle . \rangle$ whenever evaluated in the sense of
statistical mechanics.
\end{remark}
A fundamental role is played by the magnetization $m$ which we
introduce as
$$
m = \lim_{N \rightarrow \infty} m_N = \lim_{N \rightarrow \infty}
\sum_i^N \sigma_i, \ \ \langle m \rangle =  \lim_{N \rightarrow
\infty}\frac{\sum_{\sigma}m_N \exp(-\beta
H_N(\sigma))}{\sum_{\sigma} \exp(-\beta H_N(\sigma))}.
$$
Let us denote  $u_N(x,t)$  the $1$-dimesional velocity field and
$\varphi_N(x,t)$ its dynamic potential (such that $\partial_x
\varphi(x,t) = u(x,t)$). Here the label $N$ remembers us that the
analogy is made with the CW model with finite size $N$ (of course
we are interested about the thermodynamic limit of the model).

The Guerra action can be written as
\begin{equation}\label{eq:phi_N}
\varphi_N(x,t)=-\frac{1}{N}\log \sum_{\{\sigma_N\}}\exp\left( \frac{t}{2} N m^2_N +x N m_N\right)=-A_N(x,t)+O\left(\frac{1}{N}\right),
\end{equation}
\textit{i.e.}, the mean field CW pressure (up a minus sign)
\cite{barra0}, where $t$ stands for the inverse temperature
$\beta$ and $x$ takes into account the external magnetic field
$h$.

Deriving (\ref{eq:phi_N}) we get
\begin{equation}\label{eq:u_N}
u_N(x,t)=-\meanv{m_N}(x,t),
\end{equation}
the mean value of the magnetization. So our analogy is now
completed, and we can write a fluid equation as a transport
equation for $u_N(x,t)$, plus an Hamilton-Jacobi (HJ) equation for
$\varphi_N(x,t)$ and a continuity equation, defining the (purely
fictitious) density function $\rho(x,t)$.

We notice that the  Guerra action $\varphi_N(x,t)$ satisfies an HJ
equation where the potential function is the self-averaging of the
magnetization. Indeed, since we have
$$
\partial_t \varphi_N(x,t)=-\frac{1}{2}\meanv{m^2_N},
$$
and
$$
\partial^2_{x^2}
\varphi_N(x,t)=\partial_{x}u_N(x,t)=-\partial_{x}\meanv{m_N}(x,t)=-N(\meanv{m_N^2}(x,t)-\meanv{m_N}^2(x,t)),
$$
we can easily choose the external pressure for the fluid, that appears as a potential in the HJ equation, as
\begin{equation}\label{eq:V_N}
V_N(x,t)=\frac{1}{2}\Big(\meanv{m_N^2}(x,t)-\meanv{m_N}^2(x,t)\Big),
\end{equation}
and we have also
\begin{equation}\label{eq:termine-diffusivo}
-\frac{1}{2N}\partial^2_{x} \varphi_N(x,t)=V_N(x,t).
\end{equation}
Finally, computing \begin{equation}\label{fattorizzazione}
\varphi_N(x,0)=-A_N(x,0)=-\log2-\log\cosh x,
\end{equation}
we can build the differential problem for our hydrodynamical
potential $\varphi_N(x,t)$:
\begin{equation}\label{eq:HJ-N_V}
\left\{
\begin{array}{rclll}
&& \partial_t\varphi_N(x,t)+\frac{1}{2}(\partial_x\varphi_N(x,t))^2+V_N(x,t)= 0 &\:&\mbox{in }{\mathbb R}\times(0,+\infty)\\
&& \varphi_N(x,0)=-\log2-\log\cosh x&\:&\mbox{on  }{\mathbb
R}\times \{t=0\}.
\end{array}
\right.
\end{equation}
\begin{remark}
We stress that by choosing as a boundary a general point on $x$
but $t=0$ (as we did in eq.(\ref{fattorizzazione})), we implicitly
 skipped the evaluation of the two body interaction which is,
 usually, the hard core of the statistical mechanics calculations
 as the one body problem trivially factorizes.
\end{remark}
Eq. (\ref{eq:HJ-N_V}) is  just the Hamilton-Jacobi equation for a
mechanical $1$-dimensional system, with time-dependent
interactions. We can write it in a more suggestive way, for
exalting our hydrodynamical analogy. Indeed, bearing in mind
(\ref{eq:termine-diffusivo}), we have
\begin{equation}\label{eq:HJ-N-diff}
\left\{
\begin{array}{rclll}
&& \partial_t\varphi_N(x,t)+\frac{1}{2}(\partial_x\varphi_N(x,t))^2-\frac{1}{2N}\partial^2_{x}\varphi_N(x,t)= 0 &\:&\mbox{in }{\mathbb R}\times(0,+\infty)\\
&& \varphi_N(x,0)=-\log2-\log\cosh x&\:&\mbox{on  }{\mathbb
R}\times \{t=0\}.
\end{array}
\right.
\end{equation}
This equation is more interesting than the first one, for several
reasons. At first it is \textit{closed} with respect to the
unknown function\footnote{This is actually a feature of the
ferromagnets. For instance it is easily seen that it is not
trivially closed for the SK pressure because every derivation
involves different overlap combination \cite{barra1}.}.
Furthermore it has a clear physical and mathematical meaning:
Indeed the presence of a dissipative term suggests the typical
viscous fluid behavior, where friction acts against the motion.
The \textit{smallness} of this term (that appears with a factor
$N^{-1}$) acts as a mollifier for our differential problem. It may
appear even clearer by investigating the equation for $u_N(x,t)$.
Deriving with respect to $x$ eq.(\ref{eq:HJ-N-diff}) (and using
standard results about for the order of derivation) we obtain
\begin{equation}\label{eq:burgers-N}
\left\{
\begin{array}{rclll}
&& \partial_t u_N(x,t)+u_N(x,t)\partial_x u_N(x,t)-\frac{1}{2N}\partial^2_{x}u_N(x,t)= 0&\:&\mbox{in }{\mathbb R}\times(0,+\infty)\\
&& u_N(x,0) = -\tanh (x)&\:&\mbox{on  }{\mathbb R}\times \{t=0\}.
\end{array}
\right.
\end{equation}
This is a viscous Burgers' equation, i.e. a very simple
Navier-Stokes equation in one dimension. Here the mollifier term
is more incisive, since, as we will see soon, when it vanishes
(i.e. in thermodynamic limit), it induces the spontaneous
$\mathbb{Z}_2$ symmetry breaking  of statistical mechanics by
making the solution $u(x,t)$ (i.e. the magnetization) not regular
in the whole $(x,t)$ half-plane.

Lastly let us derive the continuity equation that should complete
our formal hydrodynamical analogy for the ferromagnetic model. We
stress that it does not carry any further information about the
model, as it is all contained in (\ref{eq:HJ-N-diff}) and
(\ref{eq:burgers-N})). From the continuity equation we get
\begin{eqnarray}
\partial_t \rho_N(x,t)+u_N(x,t)\partial_x \rho_N(x,t)&=&-\rho_N(x,t)\partial_x u_N(x,t)\nonumber\\
&=&\rho_N(x,t)2NV_N(x,t)\nonumber.
\end{eqnarray}
Writing
\begin{equation}\label{eq:DN}
D_N(x,t)=\partial_t+u_N(x,t)\partial_x=\frac{d}{ds},
\end{equation}
the differential operator along the stream lines, we obtain the equation for $\rho_N$
\begin{equation}\label{eq:densita-ising}
D_N(x,t)\rho_N(x,t)=2NV_N(x,t)\rho_N(x,t),
\end{equation}
solved by
\begin{equation}\label{eq:rho1}
\rho_N(x,t)=\rho_N(0,0) e^{2N\int ds V(x(s),t(s))}
\end{equation}
that is
\begin{equation}\label{eq:rho-ising}
\rho_N(x,t)=\frac{1}{2^N}\sum_{\{\sigma\}}\exp{\left[Ntm^2_N+Nxm_N\right]}=Z_N(2t,x).
\end{equation}

Resuming, mean field ferromagnets of finite size $N$ is completely
equivalent to the $1$-dimensional viscous fluid described by
equations
\begin{equation*}
\left\{
\begin{array}{rcl}
&& \partial_t u_N(x,t)+u_N(x,t)\partial_x u_N(x,t)-\frac{1}{2N}\partial^2_{x^2}u_N(x,t)= 0\\
&& D_N(x,t)\rho_N(x,t)=2NV_N(x,t)\rho_N(x,t),
\end{array}
\right.
\end{equation*}
and in thermodynamic limit, to an Eulerian fluid, such that
\begin{equation*}
\left\{
\begin{array}{rcl}
&& \partial_t u(x,t)+u(x,t)\partial_x u(x,t)= 0\\
&& \rho(x,t)^{-1} D(x,t)\rho(x,t) =0.
\end{array}
\right.
\end{equation*}
We would like now to link the finite dimensional model with its
thermodynamic limit, \textit{i.e.} the viscous fluid with the
inviscid one. It is consequently useful to study the free problem
\begin{equation}\label{eq:HJ-TL}
\left\{
\begin{array}{rclll}
&& \partial_t\varphi(x,t)+\frac{1}{2}(\partial_x\varphi(x,t))^2= 0 &\:&\mbox{in }{\mathbb R}\times(0,+\infty)\\
&& \varphi(x,0)=-\log2-\log\cosh x&\:&\mbox{on }{\mathbb
R}\times \{t=0\},
\end{array}
\right.
\end{equation}
and
\begin{equation}\label{eq:burgers-TL}
\left\{
\begin{array}{rclll}
&& \partial_t u(x,t)+u(x,t)\partial_x u(x,t)= 0&\:&\mbox{in }{\mathbb R}\times(0,+\infty)\\
&& u(x,0) = -\tanh x&\:&\mbox{on }{\mathbb R}\times \{t=0\}.
\end{array}
\right.
\end{equation}
With this purpose we can use Theorem \ref{th:lax}.
\begin{remark}
We stress that via Theorem (\ref{th:lax}) changing the boundary
condition is equivalent to modify the nature of the spin variables
in the ferromagnetic model. Since the condition on $H$ is
Lipschitz-continuity, such a theorem is valid for every distribution
of spin variables with compact support, but not for example for Gaussian
ones (at least trivially). We let for future works further investigations
\cite{future}. Hereafter anyway we will deal with only dichotomic
variables.
\end{remark}
With $h(y)=-\log2-\log\cosh y$, $y=x-tu(x,t)$ (given by (\ref{eq:u})), we find
$$
\varphi(x,t)=\frac{t}{2}u(x,t)^2-\log2-\log\cosh \left(x-tu(x,t)\right),
$$
and bearing in mind $\varphi=-A$ and $u=-\meanv{m}$, by setting
$t=\beta$ and $x=h$, we gain the usual free energy for mean field
ferromagnet
$$
f(\beta,h)=-\frac{1}{\beta}A(\beta,h)=\frac{1}{\beta}\varphi(\beta,h)=\frac{1}{\beta}\left\{ \frac{\beta \meanv{m}^2}{2}-\log\cosh\beta\left(h+\meanv{m}\right)-\log2 \right\},
$$
where of course $\meanv{m}$ is the limiting value for the
magnetization, as we are going to show. We only have to prove
convergence for differential problems (\ref{eq:HJ-N-diff}) and
(\ref{eq:burgers-N}) to the free ones, respectively
(\ref{eq:HJ-TL}) and (\ref{eq:burgers-TL}). Let us start with the
former by stating the following

\begin{teorema}\label{th:lungo}
The function
\begin{equation}\label{eq:int-gauss-ising}
\varphi_N(x,t)=-\frac{1}{N}\log\left[\sqrt{\frac{N}{t}}\int_{-\infty}^{+\infty}\frac{dy}{\sqrt{2\pi}}\exp-N\left(\frac{(x-y)^2}{2t}-\log2-\log\cosh y \right) \right]
\end{equation}
solves the differential problem (\ref{eq:HJ-TL-free}), and it is
\begin{equation}
|\varphi_N(x,t)-\varphi(x,t)|\leq O(\frac{1}{N}).
\end{equation}
\end{teorema}

\textbf{Proof} In order to find a  solution of (\ref{eq:HJ-TL}),
we put\footnote{This is usually known as the Cole-Hopf transform
\cite{evans}.}
$$
\phi_N(x,t)=e^{-N\varphi_N(x,t)}.
$$

After a few calculations
\begin{eqnarray}
\partial_t \phi_N(x,t)&=&\frac{1}{2}N\phi_N(x,t)(\partial_x\varphi_N(x,t))^2-\frac{1}{2}\phi_N(x,t)\partial^2_{x^2}\varphi_N(x,t)\nonumber\\
&=&\frac{1}{2N}\partial^2_{x^2}\phi_N(x,t), \nonumber\\
\end{eqnarray}
we see that $\phi(x,t)$ solves the heat equation with conductivity
$\frac{1}{2N}$ (and a suitable boundary condition):
\begin{equation}\label{eq:calore}
\left\{
\begin{array}{rclll}
&& \partial_t \phi_N(x,t)-\frac{1}{2N}\partial^2_{x^2}\phi_N(x,t)= 0&\:&\mbox{in }{\mathbb R}\times(0,+\infty)\\
&& \phi_N(x,0) = 2^{-N}\cosh^{-N}x&\:&\mbox{on }{\mathbb R}\times \{t=0\}.
\end{array}
\right.
\end{equation}
The unique bounded solution of (\ref{eq:calore}) is
$$
\phi_N(x,t)=\sqrt{\frac{N}{t}}\int_{-\infty}^{+\infty}\frac{dy}{\sqrt{2\pi}}\exp\left(-N\Big(\frac{(x-y)^2}{2t}-\log2-\log\cosh
y \Big)\right)
$$
and, bearing in mind $\varphi_N=-\frac{1}{N}\log \phi_N$, we have

$$
\varphi_N(x,t)=-\frac{1}{N}\log\left[\sqrt{\frac{N}{t}}\int_{-\infty}^{+\infty}\frac{dy}{\sqrt{2\pi}}\exp
\left( -N\Big(\frac{(x-y)^2}{2t}-\log2-\log\cosh y \Big)
\right)\right].
$$
We notice that, since the uniqueness of the minimum of the
function in the exponent (allowed by Theorem (\ref{th:lax})), we
easily get $\varphi_N\to\varphi$ when $N\to\infty$.

Finally bounds on the error can be made via standard techniques.
$\Box$

We must now prove an analogue result for the velocity field
$u(x,t)$. Since the equations for $\varphi(x,t)$ and $u(x,t)$ are
trivially related by a derivation, it is clear that $u_N(x,t) \to
u(x,t)$ uniformly in the thermodynamic limit. Anyway for the sake
of completeness (and as a guide for testing other models) we state
the following

\begin{teorema}\label{th:analogo-u}
The function
\begin{equation}\label{eq:int-gauss-ising-u}
u_N(x,t)=\frac{\int_{-\infty}^{+\infty}\frac{dy}{\sqrt{2\pi}}\frac{x-y}{t}\exp\Big(-N\left(\frac{(x-y)^2}{2t}-\log2-\log\cosh
y
\right)\Big)}{\int_{-\infty}^{+\infty}\frac{dy}{\sqrt{2\pi}}\exp\Big(-N\left(\frac{(x-y)^2}{2t}-\log2-\log\cosh
y \right)\Big)}
\end{equation}
solves the differential problem (\ref{eq:burgers-TL}) and it is
\begin{equation}
|u_N(x,t)-u(x,t)|\leq O\left(\frac{1}{\sqrt{N}}\right).
\end{equation}

\end{teorema}

\textbf{Proof}
The (\ref{eq:int-gauss-ising-u}) is easily obtained by direct derivation of  $\varphi_N$ in (\ref{eq:int-gauss-ising}).

Again the bound on the error is made via standard techniques.
$\Box$

Finally we can state the subsequent

\begin{corollario}\label{cor:V}
It is $V_N(x,t)\leq O(\frac{1}{N})$ a. e..
\end{corollario}

\textbf{Proof}
For the two previous theorems we have
$$
\varphi_N(x,t) =\varphi(x,t) + O(\frac{1}{N}),
$$
thus
$$
\partial_t\varphi_N=\partial_t\varphi + O(\frac{1}{N})
$$
and
$$
(\partial_x\varphi_N)^2=(\partial_x\varphi)^2 + O(\frac{1}{N}),
$$
and therefore, using the Hamilon-Jacobi equation (\ref{eq:HJ-N_V}) for $\varphi_N$, we find
$$
\partial_t\varphi+\frac{1}{2}(\partial_x\varphi)^2 + O(\frac{1}{N})+V_N=0,
$$
that implies the thesis.$\Box$

What we meant for ''a.e." is actually the whole $(x,t)$ positive
half-plane, but the line defined by $(x=0,\, t>1)$ as will be well
explained in the next section.

\subsection*{Shock waves and spontaneous symmetry breaking}

In this section we study more deeply the properties of equation (\ref{eq:burgers-TL}).
This is an inviscid  Burgers' equation, and again we can have a representation of solutions as characteristics
\cite{evans}. We get
\begin{equation}\label{eq:self-cons}
u(x,t)=-\tanh(x-u(x,t)t)
\end{equation}
\textit{i.e} the well known self consistence relation for the CW
model, with trajectories (parameterized by $s\in\mathbb{R}$)
\begin{equation}\label{eq:traiettorie-generali}
\left\{
\begin{array}{rrl}
t&=&s\\
x&=&x_0-s\tanh x_0.
\end{array}
\right.
\end{equation}
We can immediately state the subsequent
\begin{proposizione}\label{pr:shock}
In the region of the plane $(x,t)$, defined by
$$
 x\geq-\sqrt{t(t-1)}+\arctanh{\left(\sqrt{\frac{t-1}{t}} \right)}\quad\mbox{for }x_0\geq0
$$
and
$$
 x\leq-\sqrt{t(t-1)}+\arctanh{\left(\sqrt{\frac{t-1}{t}} \right)}\quad\mbox{for }x_0\leq0
$$
trajectories (\ref{eq:traiettorie-generali}) have no intersection points.
\end{proposizione}
\begin{remark}
This last statement defines the onset of ergodicity breaking in
the statistical mechanics of the CW model.
\end{remark}

\textbf{Proof} Set for instance $x_0\geq0$.

 Once fixed
$s=\bar{s}$ let us investigate the position at time $\bar{s}$ as a
function of the starting point $x_0$. We have
$$
x(x_0)=x_0-\bar{s}\tanh x_0.
$$
If $x(x_0)$ is monotone with respect to $x_0$, then $\forall
x_0\in\mathbb{R}\;\exists! x(t)$, \textit{i.e} for every starting
point there is an unique position at time $t$. In other words, two
trajectories born in different points of the boundary cannot, at
the same time, assume the same position (do not intersect). Hence
we have
$$
x^{\prime}(x_0)=1-\bar{s}(1-\tanh^2 x_0)\geq0\;\forall x_0,
$$
only if
$$
\bar{s}\leq\frac{1}{1-\tanh^2 x_0},
$$
as $1-\tanh^2x_0$ always belongs to $[0,1]$. The last formula implies
$$
x_0 \geq \arctanh{\left( \sqrt{\frac{t-1}{t}} \right)},
$$
and bearing in mind the form of trajectories
(\ref{eq:traiettorie-generali}) we get
\begin{equation}\label{eq:linea-critica} x \geq \arctanh{\left(
\sqrt{\frac{t-1}{t}} \right)}- \sqrt{t(t-1)}. \end{equation}

The proof is analogue for $x_0\leq0$. $\Box$
\newline
\newline
We must notice that the previous proposition gives  the region of
the $(x,t)$ plane in which the invertibility of the motion fails.
On the other hand, every trajectory has its end point at the
intersection with the $x$-axes, or are all merged in a unique
line, that is $(x=0,\,t>1)$.

More rigorously, the curve $(x=0,t>1)$ is a
discontinuity line for our solution, since it is easily seen that
every point of such a line is an intersection point of the
trajectories (\ref{eq:traiettorie-generali}). Also we can get by (\ref{eq:self-cons}) with direct calculation
\begin{equation}\label{eq:deriv-u-x}
\partial_x u(x,t)=-\frac{1-u^2}{1+t(1-u^2)}<0,
\end{equation}
\textit{i.e.} the velocity field is strictly decreasing along $x$
direction\footnote{This is a particular case of a more general
property of the Lax-Oleink solution \cite{lax}, named
\textit{entropy condition}, that ensures  $u(x,t)$  never
increases along $x$. We won't give the general form, that is
redundant in this contest, but can be very useful in studying
generalized ferromagnet \cite{future}.}.
\newline
Now we name $u_+$ the limiting value from positive $x$, and $u_-$
the one from negative $x$, and state the following
\begin{proposizione}
It is $0<u_-=-u_+<0$ for a.e. $t>1$.
\end{proposizione}
\textbf{Proof}  The curve of discontinuity can be parameterized as
$$
\left\{
\begin{array}{rrl}
t&>&1\\
x&=&0,
\end{array}
\right.
$$
so has zero speed. We have that $\forall \; t\geq0$ does exist a
neighbors $I$ of $(x=0)$ such that $u(x,t)$ is smooth on $I$.
Thus, since we know that our $u(x,t)$ is an integral solution, we
can use Rankine-Huginiot condition \cite{evans} to state
$$
u_+^2=u_-^2.
$$
Since for (\ref{eq:deriv-u-x}) it has to be $u_+<u_-$ the assert is proven.$\Box$
\begin{remark}
We stress that the relation $u_+^2=u_-^2$, in this context,
mirrors the spin-flip symmetry shared by the two minima of the CW
model in the broken ergodicity phase, i.e. $|+\langle m \rangle| =
|-\langle m \rangle|$.
\end{remark}
It follows that $(x=0,t>1)$ is a shock line for the Burgers'
equation (\ref{eq:burgers-TL}).

On the other hand, of course, $x=0$ is an equilibrium point for
the system, since we have that both position and velocity are
zero.
\begin{remark}
This property is translated in statistical mechanics to the
trivial case of CW model without neither a vanishing external
field, such that spontaneous magnetization can never happen.
\end{remark}
We can use it for exploring the well known mechanism of
spontaneous symmetry breaking. With this purpose, let us move on a
family of straight lines of equation
$$
x=\epsilon t-\epsilon.
$$
We have infinitely many lines, all converging in $(0,1)$, that
intersect the $x$-axes in $-\epsilon$. Let us choose for example
$\epsilon>0$, and perform the limit of $u(x,t)$ on the shock line
taking the value of $u(x,t)$ by these, and then sending
$\epsilon\to0$. Since $-\epsilon$ is negative, the intersection
point with $t=0$ is approaching $0$ from the left ($x^-$),
meanwhile the limit of $u$ is taken from the right ($u_+$). In the
same way we have that when the intersection point approach to zero
from right ($x^+$), the limiting value of u is taken from left
($u_-$).
\begin{remark}
In our analogy with statistical mechanics one can make the
substitution $u(x,t)=-\meanv{m}(h,\beta)$, and $t=\beta$,
$x=h\beta$, getting the spontaneous symmetry breaking mechanism,
in such a way that $\lim_{h\to0^{\pm}}\langle m \rangle (h,\beta) =
m^{\pm}$.
\end{remark}

\subsection*{Conservation laws}

We can rewrite the (\ref{eq:HJ-N_V}) from a mechanical point of view as
$$
\partial_t \varphi_N(x,t)+H_N(\partial_x \varphi_N(x,t),x,t)=0
$$
and the Hamiltonian function reads off as \footnote{here we name
$p$ our velocity $u$, \textit{i.e.} the velocity field coincides
with the generalized time dependent momentum}
\begin{equation}
H_N(\partial_x \varphi_N(x,t),x,t)=\frac{p^2(x,t)}{2}+V_N(x,t).
\end{equation}
Hamilton equations are nothing but characteristics of equation
(\ref{eq:HJ-N_V}):
\begin{equation}\label{eq:hamilton}
\left\{
\begin{array}{rcl}
 \dot{x}&=&u_N(x,t)\\
 \dot{t}&=&1\\
 \dot{p}&=&-u_N(x,t)\partial_x u_N(x,t)-\partial_x V_N(x,t)\\
 \dot{E}&=&-u_N(x,t)\partial_x\left(\partial_t\varphi_N(x,t)\right)-\partial_t
 V_N(x,t),
\end{array}\right.
\end{equation}
the latter two equations express the conservation laws for
momentum and energy for our system, and can be written in form of
streaming equations as
\begin{equation*}
\left\{
\begin{array}{rcl}
D_N u_N(x,t)&=&-\partial_x V_N(x,t)\\
D_N (\partial_t\varphi_N(x,t)&=&-\partial_t V_N(x,t).
 \end{array}
\right.
\end{equation*}
Since in thermodynamic limit the system approaches a free one,
bearing in mind that $u_N(x,t)=-\meanv{m_N}$ and
$\partial_t\varphi_N(x,t)=-\frac{1}{2}\meanv{m_N^2}$, so
$D_N=\partial_t-\meanv{m_N}\partial_x$, for $N\to\infty$ we
conclude
\begin{equation}\label{eq:cons-ising}
\left\{
\begin{array}{rcl}
D_N \meanv{m_N}&=&0\\
D_N \meanv{m^2_N}&=&0,
 \end{array}
\right.
\end{equation}
\textit{i.e.}
\begin{equation}\label{eq:cons-ising}
\left\{
\begin{array}{rcl}
\meanv{m^3_N}-3\meanv{m_N}\meanv{m^2_N}+2\meanv{m_N}^3&=&O(\frac{1}{N})\\
(\meanv{m^4}-\meanv{m^2}^2)-2\meanv{m}\meanv{m^3}+2\meanv{m}^2\meanv{m^2}&=&O(\frac{1}{N}).
\end{array}
\right.
\end{equation}
We have from Corollary \ref{cor:V} that
$\meanv{m^2}=\meanv{m}^2+O(\frac{1}{N})$ everywhere but on the
line $(x=0, t>1)$, where anyway $\meanv{m}=0$.  It is possible to
write down a relation that follows from energy conservation: where
the potential vanishes, using momentum conservation, giving
$\meanv{m^3}=\meanv{m}^3+O(\frac{1}{N})$, we get
$$
\meanv{m^4}-\meanv{m^2}^2=O(\frac{1}{N}).
$$
Otherwise when the potential is different from
zero\footnote{Anyway it is a zero measure set.} we have
$\meanv{m}=0$, thus the previous formula is still valid, and it
holds in all the $(x,t)$ half-plane.
\begin{remark}
This is of course a Ghirlanda-Guerra relation \cite{GG} for the CW
model (i.e. it expresses self-averaging of the internal energy
density). As a counterpart, the bare momentum conservation implies
the first Aizenman-Contucci \cite{AC} relation for $\meanv{m^3}$.
\end{remark}
\begin{remark}
It is interesting to remark that the orbits of the N\"other groups
of the theory coincide with the streaming lines of our fluid, and
conservation laws along these lines give well known identities in
the statistical mechanics of the model.
\end{remark}

\section*{The Replica Symmetric phase of the Sherringon-Kirkpatrick model}

Despite the main goal when dealing with complex systems is a clear
scenario of the {\itshape Broken Replica Phase}, which, in our
languages translates into solving viscous problems as $V_N(x,t)
\neq 0$ (and it is posted to future investigations), a detailed
analysis of the replica symmetric regime is however immediate
within this framework, as pioneered in \cite{sum-rules}.
\newline
\newline
The Sherrington-Kirkpatrick Hamiltonian is given by
$$
H_N=-\frac{1}{\sqrt{N}}\sum_{(i,j)}J_{ij}\sigma_i\sigma_j+h\sum_i\sigma_i,
$$
where $J_{ij}$ are \textit{i.i.d} centered Gaussian variables,
with $\mathbb{E}[J_{ij}]=0$ and $\mathbb{E}[J^2_{ij}]=1$.

Following \cite{sum-rules} we introduce the partition function
$$
Z_N(x,t)=\sum_{\{\sigma\}}\exp\left(\sqrt{\frac{t}{N}} \sum_{(i,j)}J_{ij}\sigma_i\sigma_j + \sqrt{x}\sum_iJ_i\sigma_i+\beta h\sum_i \sigma_i\right).
$$
Accordingly with the normalization factor $1/\sqrt{N}$ of the
model, we choose $\theta(a)=\sqrt{a}$;  it is important to stress
that differently to the ferromagnetic model, the cavity field with
strength $\sqrt{x}$ does not coincide with the magnetic field $h$,
that entries in the \textit{Boltzmannfaktor} as an external
parameter. Thus our results will hold for every value of $h$.

The main difference, when introducing thermodynamical quantities
(as the free energy) is in an overall average over the random
quenched couplings encoded in the interaction matrix. In this
sense the averages $\langle . \rangle$ now stand both for the
Boltzmann averages (denoted by $\omega$ hereafter when dealing
with a single set of phase space configuration, $\Omega=\omega
\times \omega \times ... \times \omega$ when dealing with several
replicas of the system) and for the averages over the coupling
(denoted by $\mathbb{E}$ hereafter), such that $\langle . \rangle
= \mathbb{E}\Omega(.)$.

The Guerra action for the SK model reads off as
\begin{equation}\label{eq:phi-SK}
\varphi_N(x,t)=2A_N-\frac{t}{2}-x.
\end{equation}

So it has, once introduced the two replica overlap as $q_{12} =
N^{-1}\sum_i^N \sigma_i^{(1)}\sigma_i^{(2)}$,
\begin{eqnarray}
\partial_t \varphi_N=2\partial_tA_N-\frac{1}{2}&=&-\frac{1}{2}\meanv{q_{12}^2}\label{eq:der-SK-t}\\
\partial_x \varphi_N=2\partial_xA_N-1&=&-\meanv{q_{12}}\label{eq:der-SK-x}.
\end{eqnarray}
Mirroring the mean field ferromagnet, also in this glass model the
interaction factorizes at $t=0$, and, once set
$\mathbb{E}_g=\frac{1}{\sqrt{2\pi}}\int_{-\infty}^{+\infty}dg
e^{-\frac{g^2}{2}}$, we have
$$
\varphi_N(x,0)=2A^{SK}_N(x,0)-x=2\log2+2\mathbb{E}_g\log\cosh(\beta h +g\sqrt{x})-x.
$$
The last formula, together with (\ref{eq:der-SK-t}, \ref{eq:der-SK-x}) allows to build the HJ equation for $\varphi_N(x,t)$
\begin{equation}\label{eq:HJ-SK}
\left\{
\begin{array}{rclll}
&& \partial_t\varphi_N(x,t)+\frac{1}{2}(\partial_x\varphi_N(x,t))^2+V_N(x,t)= 0 &\:&\mbox{in }{\mathbb R}\times(0,+\infty)\\
&& \varphi_N(x,0)=2\log2+2\mathbb{E}_g\log\cosh(\beta h +g\sqrt{x})-x&\:&\mbox{on }{\mathbb
R}\times \{t=0\},
\end{array}
\right.
\end{equation}
with
\begin{equation}
V_N(x,t)=\frac{1}{2}\left(\meanv{q^2_{12}}-\meanv{q_{12}}^2\right).
\end{equation}

In complete generality, this is an equation more complicated than
the ferromagnetic one: Reflecting the complex structure of the RSB
phase, the closure of the equation can be obtained only via
cumulant expansions of the overlaps in terms of higher order
correlation functions \cite{barra1}, \textit{i.e} the potential
has no trivial expression in terms of $\varphi_N$ derivatives. We
will study this equation in the Replica Symmetric phase, that is
where, in the $(x,t,h)$ domain, $\lim_NV_N=0$.

The velocity field, accordingly with  (\ref{eq:der-SK-x}), is
$$
u_N(x,t)=-\meanv{q_{12}}(x,t)
$$
and satisfies the transport equation
\begin{equation}\label{eq:burgers-SK}
\left\{
\begin{array}{rclll}
&& \partial_t u_N(x,t)+u_N(x,t)\partial_x u_N(x,t)+\partial_{x}V_N(x,t)= 0&\:&\mbox{in }{\mathbb R}\times(0,+\infty)\\
&& u_N(x,0) = -\mathbb{E}_g\tanh^2 (\beta h+g\sqrt{x})&\:&\mbox{on  }{\mathbb R}\times \{t=0\}.
\end{array}
\right.
\end{equation}
\begin{remark}
We stress that naturally in our approach the hyperbolic tangent of
the CW model has been mapped into the squared hyperbolic tangent
in the SK case, exactly as it happens in statistical mechanics,
reflecting the role of the overlap as a proper order parameter
with respect to the magnetization.
\end{remark}
Replica symmetry apart, the characteristic trajectories of
(\ref{eq:burgers-SK}) are not in general straight lines, because
of the presence of the potential. We can give an expression for
them:
\begin{equation}\label{eq:traiettorie-SK}
\left\{
\begin{array}{rclll}
t&=&s\\
x&=&x_0-s\mathbb{E}_g\tanh^2 (\beta h+g\sqrt{x_0})-\int_0^{s}ds^{\prime}\partial_xV_N(x(s),t(s)).
\end{array}
\right.
\end{equation}
and solving for $u$
\begin{equation}\label{eq:u-SK}
u_N(x,t)=-\mathbb{E}_g\tanh^2 (\beta h+g\sqrt{x_0(x,t)})-\int ds \partial_xV_N(x(s),s),
\end{equation}
where we get $x_0(x,t)$ inverting the second among (\ref{eq:traiettorie-SK}).

This is the analogous of the Guerra sum rule for the order
parameter $q$\footnote{Actually Guerra relation may be obtained
thought an integration of (\ref{eq:u-SK}).}, stating that the
difference among the true order parameter and the RS one is the
line integral of the $x$ derivative of $V_N$ along trajectories.
\newline
\newline
Reducing our attention to the RS phase of the model, we get the free HJ equation
\begin{equation}\label{eq:HJ-SK-libero}
\left\{
\begin{array}{rclll}
&& \partial_t\varphi_{RS}(x,t)+\frac{1}{2}(\partial_x\varphi_{RS}(x,t))^2= 0 &\:&\mbox{in }{\mathbb R}\times(0,+\infty)\\
&& \varphi_{RS}(x,0)=2\log2+2\mathbb{E}_g\log\cosh(\beta h +g\sqrt{x})-x&\:&\mbox{on }{\mathbb
R}\times \{t=0\},
\end{array}
\right.
\end{equation}
and Burger's equation
\begin{equation}\label{eq:burgers-SK-libero}
\left\{
\begin{array}{rclll}
&& \partial_t u_N(x,t)+u_N(x,t)\partial_x u_N(x,t)= 0&\:&\mbox{in }{\mathbb R}\times(0,+\infty)\\
&& u_N(x,0) = -\mathbb{E}_g\tanh^2 (\beta h+g\sqrt{x})&\:&\mbox{on  }{\mathbb R}\times \{t=0\}.
\end{array}
\right.
\end{equation}
We are now in perfect agreement with the hypothesis of Theorem
(\ref{th:lax}). Therefore we can write the Replica-Symmetric
Guerra action, in the thermodynamic limit, as
\begin{equation}\label{eq:phi_RS}
\varphi_{RS}(x,t)=\frac{t}{2}\left(\frac{x-y(x,t)}{t}\right)^2+\log2+\mathbb{E}_g\log\cosh(\beta h +g\sqrt{y(x,t)})-y(x,t),
\end{equation}
and, naming the velocity field of the free problem $-\bar q(x,t)$, we trivially get from (\ref{eq:u-SK}) the self-consistence equation
\begin{equation}\label{eq:qbar}
\bar q(x,t)=\mathbb{E}_g\tanh^2 (\beta h+g\sqrt{x+t\bar q(x,t)}),
\end{equation}
and the trajectories are
\begin{equation}\label{eq:traiettorie-SKRS}
\left\{
\begin{array}{rclll}
t&=&s\\
x&=&x_0-s\mathbb{E}_g\tanh^2 (\beta h+g\sqrt{x_0}).
\end{array}
\right.
\end{equation}
\begin{remark}
We stress that eq. (\ref{eq:qbar}) is exactly the self-consistent
equation for the SK model order parameter in the replica symmetric
ansatz.
\end{remark}
Furthermore the minimization point $y(x,t)$ is usually given by
$$
y(x,t)=x+t\bar q(x,t).
$$
\begin{proposizione}
For values of $t$, $x$ and $\beta h$ such that
\begin{equation}\label{eq:at}
t\mathbb{E}_g\left[\frac{1}{\cosh^4\left( \beta h + g\sqrt{x+\bar q t} \right)}\right]\leq \frac{1}{3}+\frac{2}{3}t\mathbb{E}_g\left[\frac{1}{\cosh^2\left( \beta h + g\sqrt{x+\bar q t} \right)}\right],
\end{equation}
trajectories (\ref{eq:traiettorie-SKRS}) have no intersection points.
In particular the whole region with $x\geq0$ and $t\geq0$ is included in (\ref{eq:at}).
\end{proposizione}

\begin{remark}
We notice that the (\ref{eq:at}) gives the form of the caustics for the $(x,t)$ motion, \textit{i.e.}
$$
t\mathbb{E}_g\left[\frac{1}{\cosh^4\left( \beta h + g\sqrt{x+\bar q t} \right)}\right]= \frac{1}{3}+\frac{2}{3}t\mathbb{E}_g\left[\frac{1}{\cosh^2\left( \beta h + g\sqrt{x+\bar q t} \right)}\right]
$$
and in this sense completes the theorem given in \cite{sum-rules}.
\end{remark}

\textbf{Proof} The procedure is just the same used in Proposition
\ref{pr:shock}. Starting from (\ref{eq:traiettorie-SKRS}), we put
$$
x(x_0)=x_0-t\mathbb{E}_g\tanh^2\left( \beta h + g\sqrt{x_0}\right),
$$
\textit{i.e.} the position depending by initial data, and let's
study its monotony. Given the trajectories, it is clear that,
whereas there is no intersection, $x(x_0)$ must be increasing,
thus
$$
\partial_{x_0}x(x_0)=1-t\mathbb{E}_g\partial_{x_0}\tanh^2\left( \beta h + g\sqrt{x_0}\right)\geq0,
$$
(of course we can swap derivatives and Gaussian integral). So we
have
\begin{equation}\label{eq:at-formula1}
t\mathbb{E}_g\partial_{x_0}\tanh^2\left( \beta h + g\sqrt{x_0}\right)\leq 1.
\end{equation}
Now
\begin{eqnarray}
\mathbb{E}_g\partial_{x_0}\tanh^2\left( \beta h + g\sqrt{x_0}\right)&=&\frac{1}{\sqrt{x_0}}\mathbb{E}_g\left[g \frac{\tanh\left( \beta h + g\sqrt{x_0}\right)}{\cosh^2\left( \beta h + g\sqrt{x_0}\right)} \right]\nonumber\\
&=&\frac{1}{\sqrt{x_0}}\mathbb{E}_g\left[\partial_g \frac{\tanh\left( \beta h + g\sqrt{x_0}\right)}{\cosh^2\left( \beta h + g\sqrt{x_0}\right)} \right]\nonumber\\
&=&\mathbb{E}_g\left[\frac{1}{\cosh^4\left( \beta h + g\sqrt{x_0}\right)}\right]-2\mathbb{E}_g\left[\frac{\tanh^2\left( \beta h + g\sqrt{x_0}\right)}{\cosh^2\left( \beta h + g\sqrt{x_0}\right)}\right]\nonumber\\
&=&3\mathbb{E}_g\left[\frac{1}{\cosh^4\left( \beta h + g\sqrt{x_0}\right)}\right]-2\mathbb{E}_g\left[\frac{1}{\cosh^2\left( \beta h + g\sqrt{x_0}\right)}\right]\nonumber
\end{eqnarray}
where we have used the well known formula for Gaussian expectation
$\mathbb{E}_g\left[gF(g)\right]=\mathbb{E}_g\left[\partial_g
F(g)\right]$. At this point,  putting the last expression in
(\ref{eq:at-formula1}) we gain the (\ref{eq:at}). $\Box$

We can finally give the form of the Sherrington-Kirkpatrick
solution for the pressure of the model \cite{sk1}\cite{MPV}. It is
\begin{eqnarray}
A_{RS}(\beta)&=&A_{RS}(0,\beta^2)\nonumber\\
&=&\frac{1}{2} \varphi_{RS}(0,\beta^2)+\frac{\beta^2}{4}\nonumber\\
&=&\log 2 + \mathbb{E}_g\log\cosh(\beta h +g\beta\sqrt{\bar
q})+\frac{\beta^2}{4}\left(1-\bar q\right)^2.
\end{eqnarray}

\subsection*{Conservation laws}

In the same way we did for the CW model, we can get relation among
overlap from momentum and energy conservation laws, holding in RS
regime.  It is remarkable that the vanishing, in thermodynamic
limit, of an overlap polynomial is associated to a N\"other
streaming of mechanical quantities.

With the aim of deepen this last paragraph, let us stating the
following
\begin{lemma}
Given $F(\sigma^1...\sigma^s)$ as a smooth, well behaved function
of $s$ replicas, we have
$$
D\meanv{F}=\frac{N}{2}\meanv{F\left[ \sum_{a\leq
b}^{s}q^2_{ab}-s\sum_a^s q^2_{a,s+1}+\frac{s(s+1)}{2}q^2_{s+1,s+2}
\right]}
$$
\end{lemma}
The proof of this lemma works via a long  and direct calculation,
and we will not report it here \cite{barra1}\cite{sum-rules}.
\newline
\newline
We stress that the linearity of $D$ implies all our relations
 approach zero as $O\left(1/N\right)$.

We have, in general, that conservation laws for momentum and
energy are given by the streaming equation
\begin{eqnarray}
D_N\meanv{q_{12}}&=&-\partial_xV_N(x,t)\label{eq:cons-SK-1}\\
D_N\meanv{q^2_{12}}&=&-2\partial_tV_N(x,t)\label{eq:cons-SK-2}.
\end{eqnarray}
Of course in RS phase, the right hand term of (\ref{eq:cons-SK-1})
and (\ref{eq:cons-SK-2}) vanishes when $N\to\infty$. Although such
an approach does not give any information about the way they
vanish in thermodynamic limit (we can always write it is $o(1)$),
we can write down our relation without problems, since the
presence of $D$ forces them to be at least $O(1/N)$. Explicitly we
get
\begin{eqnarray}
N\meanv{q^3_{12}-4q_{12}q^2_{23}+3q_{12}q^2_{34}}-N\meanv{q_{12}}\meanv{q^2_{12}-4q_{12}q_{23}+3q_{12}q_{34}}&=&o(1)\nonumber\\
N\meanv{q^4_{12}-4q^2_{12}q^2_{23}+3q^2_{12}q^2_{34}}-N\meanv{q_{12}}\meanv{q^3_{12}-4q_{12}q^2_{23}+3q_{12}q^2_{34}}&=&o(1),\nonumber
\end{eqnarray}
\textit{i.e.} conservation of momentum and energy along the
streaming lines of the systems (or along free trajectories
(\ref{eq:traiettorie-SKRS})) implies that in  the RS regime
\begin{eqnarray}
\meanv{q^3_{12}-4q_{12}q^2_{23}+3q_{12}q^2_{34}}_{(x,t)}-\meanv{q_{12}}_{(x,t)}\meanv{q^2_{12}-4q_{12}q_{23}+3q_{12}q_{34}}_{(x,t)}&\leq&O\left(\frac{1}{N}\right) \nonumber\\
\meanv{q^4_{12}-4q^2_{12}q^2_{23}+3q^2_{12}q^2_{34}}_{(x,t)}-\meanv{q_{12}}_{(x,t)}\meanv{q^3_{12}-4q_{12}q^2_{23}+3q_{12}q^2_{34}}_{(x,t)}&\leq&O\left(\frac{1}{N}\right).\nonumber
\end{eqnarray}
Combining the previous results, we get a third relation
\begin{equation}\label{eq:cons-combinata}
\meanv{q^4_{12}-4q^2_{12}q^2_{23}+3q^2_{12}q^2_{34}}_{(x,t)}-\meanv{q_{12}}^2\meanv{q^2_{12}-4q_{12}q_{23}+3q_{12}q_{34}}_{(x,t)}\leq
O\left(\frac{1}{N}\right),
\end{equation}
which, in particular we find physically meaningful, when setting
$x=0$ and $t=\beta^2$, because the replica symmetric assumption on
the vanishing of the potential is clearly reflected into the
overlap labels in the last identity.
\newline
\newline
If  now we neglect the magnetic field $(h=0)$, as we are in the
replica symmetric regime, the gauge symmetry holds such that the
SK Hamiltonian is left invariant under the transformation
$\sigma\to\sigma\bar\sigma$, $\bar\sigma$ being a dichotomic
variable out from the $N$-spin Boltzmann average. Matching
\cite{barra1} and \cite{sum-rules} in fact it is straightforward
to check that gauging the energy conservation we get (again we
stress that it holds only at $h=0$, and obviously  at $t=\beta^2$
e $x=0$)
$$
(1-\meanv{q^2_{12}})\meanv{q^4_{12}-4q^2_{12}q^2_{23}+3q^2_{12}q^2_{34}}\leq O\left(\frac{1}{N}\right)
$$
and consequently
$$
\meanv{q^4_{12}-4q^2_{12}q^2_{23}+3q^2_{12}q^2_{34}}\simeq
O\left(\frac{1}{N}\right),
$$
obtaining the well known relation constraining overlaps
\cite{AC}\cite{barra1}.

\section*{Conclusions and outlook}

In this work we built a self-consistent method to solve for the
thermodynamics of mean field systems, encoded by self-averaging
order parameters.

Such a method minimally relies on statistical mechanics,
essentially just on the boundary conditions of our partial
differential equations, and however, involves just straightforward
one-body problems.

Within our approach, that we tested on the Curie-Weiss prototype,
we obtained the explicit expression for the free energy as a
solution of an Hamilton-Jacobi equation defined on a $1+1$
Euclidean space time, whose velocity field obeys a suitably
defined Burger's equation in the same space.

The critical line defining ergodicity breaking is obtained as a
shock wave for a properly defined  Cauchy problem. The behavior of
the magnetization, thought of as this velocity field, both in the
ergodic and in the broken ergodicity phases have also been
obtained rigorously.

As instruments involved in our derivation, we obtained rigorously
also the existence of the thermodynamic limit for the free energy
and the self-averaging of the order parameter.

Despite the problems in relating conserved quantities and discrete
symmetries, in our continuous framework, Noether theory is
straightforwardly applicable and gives the well known
factorization of the momenta of the order parameter, as expected,
being the Curie-Weiss a mean field model.

Furthermore we applied the method even to the replica symmetric
phase of the Sherrington-Kirkpatrick model, founding full
agreement with Guerra's results and stressing other points as the
study of the caustics (which shares some similarities with the AT
line) and the study of the symmetries, which turn out to be
polynomial identities, typical of complex systems (first of all
the Aizenman-Contucci relations).

We emphasize that, actually, we believe the method working for a
large range of models (\textit{i.e.} with general interacting
variables as spherical spins, etc), several interacting spins
 as P-spin models, etc...). However, of course, it is still not
enlarged to cover the case of not self-averaging order parameters
(which is mathematically challenge even with already structured
methods).

Furthermore a certain interest should be payed trying to apply the
method to finite dimensional problems.

We plan to report soon on these topics and their possible
applications.

\section*{Acknowledgments} The authors are grateful to Francesco
Guerra for a priceless scientific interchange. Further they are
pleased to thank Pierluigi Contucci, Sandro Graffi, Isaac Perez
Castillo and Renato Luc\'a for useful discussions.
\newline
AB work is supported by the SmartLife Project (Ministry Decree
$13/03/2007$ n.$368$). GG work is supported by a Techonogical
Voucher $B$ via the Physics Department of Parma University.

\end{document}